\documentstyle[12pt,epsfig]{article}

\def\@sect#1#2#3#4#5#6[#7]#8{\ifnum #2>\c@secnumdepth
     \let\@svsec\@empty\else  
     \refstepcounter{#1}\edef\@svsec{\csname the#1number\endcsname}\fi
     \@tempskipa #5\relax
      \ifdim \@tempskipa>\z@
        \begingroup #6\relax
          \@hangfrom{\hskip #3\relax\@svsec}{\interlinepenalty \@M #8\par}%
        \endgroup
       \csname #1mark\endcsname{#7}\addcontentsline
         {toc}{#1}{\ifnum #2>\c@secnumdepth \else
                      \protect\numberline{\csname the#1\endcsname}\fi
                    #7}\else
        \def\@svsechd{#6\hskip #3\relax  
                   \@svsec #8\csname #1mark\endcsname
                      {#7}\addcontentsline
                           {toc}{#1}{\ifnum #2>\c@secnumdepth \else
                             \protect\numberline{\csname the#1\endcsname}\fi
                       #7}}\fi
     \@xsect{#5}}

\advance\textheight by \topskip
 \advance\footskip by 6mm

\renewenvironment{abstract}{\if@twocolumn \section*{ABSTRACT}\else
\begin{quote}\begin{center}%
\bf \abstractname\par
\end{center}\vskip 0.5ex\fi}
{\if@twocolumn\else\end{quote}\vskip 3ex\fi}

\def\nostrocostrutto#1\over#2{\mathrel{\mathop{\kern 0pt \rlap 
  {\raise.2ex\hbox{$#1$}}}
  \lower.9ex\hbox{\kern-.190em $#2$}}}

\newcommand{\be}{\begin{equation}}
\newcommand{\ee}{\end{equation}}
\newcommand{\ba}{\begin{eqnarray}}
\newcommand{\ea}{\end{eqnarray}}

\newcommand{\eref}[1]{(\ref{#1})}      
\newcommand{\pp}{{$\pm$}}
\newcommand{\N}{{\mathcal N}}

\catcode`@=11
\newcount\@tempcntc
\def\@citex[#1]#2{\if@filesw\immediate\write\@auxout{\string\citation{#2}}\fi
  \@tempcnta\z@\@tempcntb\m@ne\def\@citea{}\@cite{\@for\@citeb:=#2\do
    {\@ifundefined
       {b@\@citeb}{\@citeo\@tempcntb\m@ne\@citea\def\@citea{,}{\bf ?}\@warning
       {Citation `\@citeb' on page \thepage \space undefined}}%
    {\setbox\z@\hbox{\global\@tempcntc0\csname b@\@citeb\endcsname\relax}%
     \ifnum\@tempcntc=\z@ \@citeo\@tempcntb\m@ne
       \@citea\def\@citea{,}\hbox{\csname b@\@citeb\endcsname}%
     \else
      \advance\@tempcntb\@ne
      \ifnum\@tempcntb=\@tempcntc
      \else\advance\@tempcntb\m@ne\@citeo
      \@tempcnta\@tempcntc\@tempcntb\@tempcntc\fi\fi}}\@citeo}{#1}}
\def\@citeo{\ifnum\@tempcnta>\@tempcntb\else\@citea\def\@citea{,}%
  \ifnum\@tempcnta=\@tempcntb\the\@tempcnta\else
   {\advance\@tempcnta\@ne\ifnum\@tempcnta=\@tempcntb \else \def\@citea{--}\fi
    \advance\@tempcnta\m@ne\the\@tempcnta\@citea\the\@tempcntb}\fi\fi}
\catcode`@=12

\begin{document}

\noindent
\rightline{Durham DTP/96/38}
\rightline{MPI-PhT/96-28}
\rightline{April, 1996}
\vspace{1.0cm}
\begin{center}
{\Large\bf Perturbative description of particle spectra} 
\end{center}
 \begin{center}
{\Large\bf  at LEP-1.5} 
\end{center}
\vspace{0.4cm} 
\begin{center}
VALERY A. KHOZE$^{1,2,3}~$\footnote{e-mail: v.a.khoze@durham.ac.uk} \ , \ 
SERGIO LUPIA$^3$~\footnote{e-mail: lupia@mppmu.mpg.de} \ , \ 
WOLFGANG OCHS$^3$~\footnote{e-mail: wwo@mppmu.mpg.de} 
\end{center}

\begin{center}
$^1$ \ {\it Department of Physics\\
University of Durham, Durham DH1 3LE, UK}\\ 
\mbox{ }\\
$^2$ \ {\it Institute for Nuclear Physics\\
St. Petersburg, Gatchina, 188350, Russia}\\
\mbox{ }\\
$^3$ \ {\it Max-Planck-Institut f\"ur Physik \\
Werner-Heisenberg-Institut\\
F\"ohringer Ring 6, D-80805 Munich, Germany} 
\end{center}
\vspace{1.0cm}

\begin{abstract}
The recent data from LEP-1.5 on charged particle spectra are analyzed within
the analytical QCD approach. 
\end{abstract}

\newpage

\section{Introduction} 

$e^+e^-$ annihilation into hadrons proves to be a  wonderful laboratory for
detailed experimental tests of QCD. The high statistics data collected from
hadronic $Z^0$ decays in LEP-1 and SLD experiments ~\cite{ada1} allow one to
perform detailed studies of perturbative QCD and to reduce the domain of
our ignorance on the physics of confinement. 
The data have convincingly demonstrated the dominant role of the perturbative
phase of jet evolution and supported the hypothesis of local parton-hadron
duality~\cite{adkt1,dkmt2}. We have now quite successfully entered the 
stage of quantitative tests of the so-called Modified Leading Log 
Approximation (MLLA)
which allows one to calculate systematically 
the inclusive jet characteristics up to terms of relative order
$\sqrt{\alpha_s}$~\cite{dkmt2,dt,ahm1}. In some special cases next-to-MLLA
corrections are also calculated, (for reviews see~\cite{dkmt1,dkmt2,KO}). 

In November 1995 LEP was operated at centre-of-mass energies 130-140 GeV
(LEP-1.5). The first experimental data~\cite{aleph15,delphi15,l315,opal15} 
show good agreement with perturbative expectations. In particular, 
semihard QCD results have become available now. 

Further studies of the various aspects of QCD dynamics will be provided by
LEP-2 and future linear $e^+e^-$ colliders (for discussion, 
see e.g. ~\cite{vak2} and ~\cite{vak1}) as well
as by HERA and TEVATRON. It is worthwhile to mention the unique 
opportunity to compare the particle spectra measured by the
same experimental group at the same accelerator at different energies.

The aim of this paper is twofold. First of all we would like to perform the
complete analysis of the recent data on inclusive charged particle
distributions collected by LEP-1.5. 
We show that  the inclusive
charged particle spectra and mean multiplicity agree impressively well  
with the analytical QCD results. Secondly we would like to discuss 
the working tools relevant 
for the detailed treatment of the forthcoming data. 

\section{Inclusive charged particle spectrum in QCD jets} 
In the MLLA this spectrum can be obtained as solution of the appropriate
evolution equation~\cite{dkmt2,dt} in terms of two parameters,
 the QCD scale $\Lambda$ and 
the $k_\perp$  cut-off
$Q_0$ in the cascade. In the case when both parameters coincide
($Q_0=\Lambda$) one obtains the so-called limiting parton spectrum 
\cite{dkmt1,dkt5} which has been                           
 proven to be very successful in fitting the experimental data of charge
 particle production in QCD jets.

Until now, 
characteristics of an individual quark jet have been  studied in greatest
detail 
experimentally in the reaction $e^+ e^- \rightarrow$ hadrons.  An
inclusive momentum spectrum here is the sum of two $q$-jet
distributions.  In terms of the limiting spectrum
one obtains for the distribution in the variable $\xi = \log
1/x$ with $x = 2 E_h/\sqrt{s}$  
\begin{equation}
\frac{1}{\sigma} \; \frac{d \sigma^h}{d \xi} \; 
    = \; 2 K^h  D^{\lim}_q(\xi,Y)
\label{spectrum}
\end{equation}
 where  $K^h$ is the hadronization constant,  
 $\sqrt{s}$ the total $cms$ energy and $Y = \log (\sqrt{s}/2 \Lambda)$. 
The limiting spectrum is readily
 given using an integral representation for the confluent hypergeometric
 function~\cite{dt,dkmt2}: 
\begin{eqnarray}
\label{LS}
D^{\lim}_q(\xi,Y) & = & \frac{4 C_F}{b} \: \Gamma (B) \;
\int_{-
\frac{\pi}{2}}^{\frac{\pi}{2}} \; \frac{d \ell}{\pi} \: e^{- B
\alpha} \; \left [ \frac{\cosh \: \alpha + (1 - 2 \zeta) \sinh
\: \alpha}{\frac{4 N_C}{b} \; Y \; \frac{\alpha}{\sinh \:
\alpha}}
\right ]^{B/2} \nonumber \\
& & \\
& & \times \; I_B \; \left ( \sqrt{ \frac{16 N_C}{b} \; Y \;
\frac{\alpha}{\sinh \: \alpha} \; [\cosh \: \alpha + (1 - 2
\zeta) \: \sinh \: \alpha ]} \right) . \nonumber
\end{eqnarray}
\noindent Here $\alpha = \alpha_0 + i \ell$ and $\alpha_0$ is
determined by $\tanh \: \alpha_0 = 2 \zeta - 1$ with $\zeta =
1 - \frac{\xi}{Y}$.  $I_B$ is the modified Bessel function of
order $B$, where $B =  a/b$, $a = 11 N_C/3 + 2 n_f/ 3 N_C^2$, 
$b = (11 N_C - 2 n_f)/3$, with $n_f$ the number of flavours and 
$C_F = (N_C^2 -1)/2N_C$ = 4/3. 

The analysis of charged particle spectra using this distribution 
\cite{opal,dkt9,lo} yields values for the  
effective scale parameter  $\Lambda \equiv \Lambda_{ch}$
in the range 
 $\Lambda_{ch} \simeq 250 \div 270$ MeV.  
If both parameters $Q_0$ and $\Lambda$ are kept free in the fit one 
recovers the limiting spectrum with $Q_0=\Lambda$ as best solution 
\cite{lo}.

It proves to be very convenient (see e.g.\ ~\cite{fw,dkt5,lo}) to analyze 
inclusive particle spectra in terms of the normalized moments
\begin{equation}
  \xi_q \equiv <\xi^q>
    = \frac{1}{\bar \N_E} \int d\xi \xi^q D(\xi)
\end{equation}
 where $\bar \N_E$ is the mean
multiplicity in the jet, the integral of the spectrum.  These moments 
characterize the shape of the distribution and are independent of
normalization uncertainties. The theoretical predictions for the moments 
from the Limiting Spectrum are
determined by only one free parameter $\Lambda_{ch}$. 
Also one defines the cumulant moments
$K_q (Y, \lambda)$ or the reduced cumulants $k_q \equiv
K_q/\sigma^q$ ~\cite{so} which  are related by
\begin{eqnarray}
K_1 & \equiv & \overline{\xi} \; \equiv \; \xi_1 \nonumber\\
K_2 & \equiv & \sigma^2 \; = \; < (\xi - \overline{\xi})^2
>, \nonumber \\
K_3 & \equiv & s \sigma^3 = < (\xi - \overline{\xi})^3 >, \nonumber \\
K_4 & \equiv & k \sigma^4 = 
< (\xi - \overline{\xi})^4 > \: - \: 3 \sigma^4 
\label{defmom}
\end{eqnarray}
\noindent where the third and fourth reduced cumulant moments are 
the skewness $s$ and the kurtosis $k$ of the
distribution. 
If the higher-order cumulants $(q > 2)$ are sufficiently
small, one can reconstruct the $\xi$-distribution from the
distorted Gaussian formula,  see ~\cite{fw}. 

The cumulant moments can be obtained from~\cite{fw} 
\begin{equation}
K_q (Y, \lambda) \; =  \int_0^{Y} \;
dy \: \left . \left ( - \frac{\partial}{\partial \omega} \right)^p 
\; \gamma_\omega (\alpha_s (y)) \right |_{\omega = 0} 
\label{moments}
\end{equation}
where $\gamma_{\omega}(\alpha_s(y))$ denotes the anomalous dimension which
governs the energy evolution of the Laplace transform $D_{\omega}(Y)$ of the
$\xi$-distribution $D(\xi,Y)$. 
Equation~\eref{moments} shows the direct dependence of the
moments on $\alpha_s (Y)$ and thereby implicitly on $n_f$.

The analytical procedure of ~\cite{dkt5} allows one to arrive at the 
following expression for the moments: 
\be 
\bar \N_{LS} = \Gamma (B) \left( \frac{z}{2} \right)^{1-B} I_{B+1}(z) 
\label{normls}
\ee
\begin{equation}
\frac{<\xi^q>}{Y^q} = P_0^{(q)}(B+1,B+2,z) + \frac{2}{z}
\frac{I_{B+2}(z)}{I_{B+1}(z)} P_1^{(q)}(B+1,B+2,z)
\label{lsmoments}
\end{equation}
where the 
parameter $B$ is introduced above and the variable $z$ is given by 
$z  \equiv \sqrt{ 16 N_c Y / b }$; 
$P_0^{(q)}$ and $P_1^{(q)}$ are polynomials
of order $2(q-1)$ in $z$. Explicit results for the full expressions  
 for $q < 3$ can be found in ~\cite{dkt5} and for $q$ = 3,4 in ~\cite{lo1}.

In the calculations, the partons were taken 
massless and the  $p_t$ cut-off $Q_0 \equiv \Lambda_{ch}$  was introduced 
for regularization; 
on the other hand, the observable hadrons are massive. 
One can make the simple 
assumption that the cut-off $Q_0$ is related to the masses of
hadrons. As a first stage in the discussion of charged particle spectra, one
can relate $Q_0$ to an effective hadron mass, then for both 
partons and hadrons  $E \ge Q_0$. 
A consistent kinematical behaviour of parton and hadron spectra can be 
obtained through the relation~\cite{lo}: 
\be
E_h \frac{dn(\xi_E)}{dp_h}=E_p\frac{dn(\xi_E)}{dp_p} \quad ;  \quad 
E_h = E_p \ge Q_0
\label{dual}
\ee
at the same energies $E_i$ or $\xi_{E_i} = \log (\sqrt{s}/2 E_i)$ 
(not momentum!), where 
$E_h = \sqrt{p_h^2+Q_0^2}$ and $E_p = p_p$.  With this choice both spectra
vanish linearly for  $E \to Q_0$ or $\xi_E \to Y$.

\section{Comparison with data}

LEP-1.5 opens a unique opportunity to compare the high energy data 
collected at the same accelerator by the same Collaboration\footnote{at 
lower energies such a comparison has been performed by TASSO Collaboration 
at PETRA~\cite{tasso}.}. 
Later on such a comparison will be continued at LEP-2. 

\noindent {\it Momentum spectrum}

Figure~1a shows experimental data on charged particle inclusive momentum
distributions, $dn/d\xi_p$ as a function of 
$\xi_p =  \log (\sqrt{s}/2 p)$, obtained by the 
OPAL Collaboration at $cms$ energies $\sqrt{s}$ = 91.2 GeV 
(LEP-1)~\cite{opal}  and $\sqrt{s}$ = 133 GeV (LEP-1.5)~\cite{opal15}. 
Theoretical predictions of the
Limiting Spectrum~\eref{LS} at the same $cms$ energies are also shown; 
at both $cms$ energies, 
the $\Lambda_{ch}$ parameter has been taken equal to 270 MeV, as 
suggested by the moment analysis performed in ~\cite{lo}, whereas the free 
overall normalization factor has been fixed to the value
 $K^h$ = 1.31\footnote{OPAL Collaboration~\cite{opal} found a best value of
$\Lambda_{ch}$ = 253 MeV at LEP-1 with $K_h$ = 1.28 from a fit to the shape,
whereas $\Lambda_{ch}$ = 263$\pm$ 4 MeV was obtained from a fit to the peak
position by using the asymptotic formula\cite{opal15}. 
The value for
$\Lambda_{ch}$ = 270 MeV 
has been chosen in \cite{lo} to reproduce the 
moments over a large energy interval. With a restriction to the LEP-1 data, a
smaller value of $\Lambda_{ch}$ would fit slightly better.}. 
In view of the forthcoming run at LEP-2, theoretical predictions 
for $\sqrt{s}$ = 200 GeV  with the same choice of parameters are also shown. 
In Figure~1b 
data on inclusive momentum distributions extracted at LEP-1.5 $cms$ energy  by
different experimental Collaborations, i.e., ALEPH~\cite{aleph15}, 
DELPHI~\cite{delphi15}\footnote{DELPHI published the charged particle 
inclusive  energy distribution obtained by assigning  to all particles the 
pion mass. The inclusive momentum distribution
has been extracted from experimental data by using the same assignment to the
mass of all particles.} and OPAL~\cite{opal15} are compared to the same
theoretical predictions as in Figure~1a. 
The Limiting Spectrum predictions with this choice of parameters reproduce 
very well the experimental shape around the peak at LEP-1 $cms$ energy; 
the agreement at LEP-1.5 is reasonable, even though deviations are 
visible in the region  around the maximum. However, this effect could 
be well attributed to statistical
and/or systematic uncertainties, as suggested by the large spread of
experimental points shown in Figure~1b. 

Looking first at the large $x_p$ (small $\xi_p$) region, 
the experimental data can be well described at both $cms$ energies by
the Limiting Spectrum prediction~\eref{spectrum}, 
contrary to the simple distorted Gaussian
parametrization (not shown). The observed agreement of the Limiting Spectrum
prediction  in a kinematical region 
where the MLLA approach is not expected {\it a priori} to be valid is  
due to the fact that the approximate expression for the anomalous
dimension derived within MLLA at small $x_p$ turns out to mimic reasonably 
well the expression valid at large $x_p$ (see~\cite{dkt7,dkt9,kdt} 
for more details). 

In the small $x_p$ (large $\xi_p$) region, data show a tail which is not
well reproduced by theoretical predictions taking $\xi_p$ as an argument
in~\eref{LS}. This discrepancy is actually 
due to kinematical effects~\cite{dkt9,lo}. As previously discussed, 
theoretical predictions vanish  for
$E < Q_0$, i.e., for $\xi_E > Y$. This kinematical effect can be
 taken into account in a simple way by using the relation~\eref{dual} 
 between parton and hadron spectra.  The corresponding 
theoretical predictions for the spectrum can then be written as: 
 \be
 \frac{1}{\sigma}\frac{d\sigma^h}{d\xi_p} = 2 K^h 
 \frac{p}{E} D^{\lim}_q(\xi_E,Y)  \qquad , 
 \qquad \xi_E = \log \frac{\sqrt{s}}{2 \sqrt{s e^{- 2 \xi_p}/ 4 
 + Q_0^2}}  
\label{modified}
 \ee
 For $E \gg Q_0$ these differences vanish and $dn/d\xi_p \simeq
 dn/d\xi_E$, $\xi_E \simeq \xi_p$. 
It is seen in Figures~1a and ~1b (dashed lines) that they  
closely  follow the large $\xi$ tail of the experimental data. 
The best fit here is obtained with $K^h$ = 1.34. 
It is also worth noticing that a common  normalization factor, $K^h$, 
has been  used at both $cms$ energies; this result is consistent with 
the expectations of LPHD. 

\noindent {\it Moments} 

The moments $< \xi^q>$ and the cumulants $K_q$ 
are determined from the spectra $E dn/dp$ as a function of  $\xi_E$; 
the average multiplicity $\bar \N_E$ is obtained as the 
integral over $\xi_E$ of the full spectrum $E dn/dp$. 
For the unmeasured interval near $\xi_E \simeq Y$ 
(small momenta) a contribution was found by linear extrapolation as in
Notice that $\bar \N_E$ coincides with the usual particle multiplicity, 
$\bar \N$, the integral over $\xi_p$, at asymptotic energies. At LEP-1 
(LEP-1.5) the difference is about 10\% (8\%)\footnote{Good fits have 
been obtained also for the standard charged particle multiplicity, $\bar
\N$~\cite{aleph15,delphi15,l315,opal15}.}. 
The cumulants up to order $q$=4 together with the corresponding 
MLLA predictions at $cms$ energy $\sqrt{s}$ = 133 GeV
are shown in Table~1. Notice that DELPHI has also 
presented data at the $Z^0$ peak from radiative events collected at LEP-1.5. 
Cumulants extracted from different experiments are consistent within
experimental uncertainties; it is important to stress that also radiative
events at the $Z^0$ peak collected at LEP-1.5 are completely consistent with
LEP-1 results. 
Errors quoted in the Table are  statistical errors only. 
Note that moments of order $q \ge$ 1 
are independent of the overall normalization, which is the main source of
systematic uncertainties. On the other hand, 
a systematic error of the order of 0.5 should be taken into account for 
the average multiplicity at LEP-1.5  energy. Also shown in the Table are 
the results~\cite{lo} obtained at LEP-1 for comparison. 
Theoretical predictions are shown for the Limiting Spectrum with 
$\Lambda_{ch}$ = 270 MeV; 
for the average multiplicity $\bar \N_E$ the first number in the Table refers
to the prediction~\eref{normls} normalized at LEP-1, the
second number to the extrapolation of the fit  
$\bar \N_E = c_1 \frac{4}{9} 2 \bar \N_{LS} + c_2$ to the energy region 3-91
GeV~\cite{lo}. 
The agreement between experimental results and theoretical predictions is
very satisfactory; 
the deviation visible in the prediction of the average
multiplicity at LEP-1.5 is well inside the errors if one correctly takes into
account systematics. 

\noindent {\it Peak position} 

An easily accessable characteristic of the $\xi$-distribution is its maximum
$\xi^*$ which has been extensively studied by the experimental groups (for a
recent review, see e.g. ~\cite{ms}). The high energy behaviour of this 
quantity for the Limiting Spectrum is predicted as~\cite{dkt5,dkt7}: 
\begin{equation}
\xi^* \; = \; Y \: \left [ \frac{1}{2} \: + \: \sqrt{
\frac{C}{Y}} \: - \: \frac{C}{Y} \right ]
\label{asy}
\end{equation}
\noindent with the constant term  given by  
$$
C \; = \; \frac{a^2}{16 \: N_C b} \; = \; 0.2915 \: (0.3513)
\;
{\rm for} \; n_f \: = \: 3 (5).
$$
Alternatively, one can compute the maximum $\xi^*$ from the Distorted 
Gaussian approximation: 
\be
\xi^* = \overline{\xi} -  \frac{1}{2} s \: \sigma 
\label{lsdist}
\ee
using the full expression~\eref{lsmoments} for the moments defined
in~\eref{defmom}. 
One neglects here the contribution of cumulants of order greater than 4
to the position of the maximum. 
The third possibility is to extract numerically the actual position of the
maximum $\xi^*$ from the Limiting Spectrum~\eref{LS}; 
the comparison between the three choices can give us some
information on the validity of the approximations performed in the first 
two cases. 
A comparison of these three possible approaches is shown in Figure~2a for
$n_f$ = 3. Notice that since we are plotting $\xi^*$ vs. $Y$, there is no
dependence of theoretical predictions on $\Lambda_{ch}$. 

One can conclude from Figure~2a 
that the energy dependence of the maximum $\xi^*$, i.e., the slope
of the curve, is very similar in the three cases; a finite shift, however, 
is present; for example the true maximum of the Limiting Spectrum $\xi^*$ can
be expressed in terms of the approximation~\eref{asy} as: 
\begin{equation}
\xi^* \; = \; Y \: \left [ \frac{1}{2} \: + \: \sqrt{
\frac{C}{Y}} \: - \: \frac{C}{Y} \right ] + 0.10 
\label{asy2}
\end{equation} 
(The numerical term actually decreases slightly from 0.11 at $\sqrt{s}$ = 10
GeV to 0.086 at 130 GeV). 

In Figure~2b we compare experimental data on
the maximum $\xi^*$ extracted from Gaussian or Distorted Gaussian
fits\footnote{We used the $\xi^*$ values from the original publications or 
from the fits\cite{ms}.} to the central region of the inclusive momentum
spectrum~\cite{tasso,mark2,tpc,cello,amy,topaz,%
aleph,delphi,l3,opal,aleph15,opal15} with the theoretical
predictions obtained extracting the maximum from the shape of the Limiting
Spectrum. We found a good agreement for $\Lambda_{ch}$ = 270 MeV, which also
describes the energy evolution of the moments. 
The best value of $\Lambda_{ch}$ which fits the 
experimental data is slightly different, depending on which theoretical
formula is beeing used. 
For instance,  by using the asymptotic formula~\eref{asy2}, 
a smaller value of $\Lambda_{ch}$ would have been better. 
The difference in the $\Lambda_{ch}$ value is of the 
order of 20-30 MeV. 

Another possible source of ambiguity is the number of active flavours to 
be used in the theoretical formula. 
This problem can be studied most easily\cite{lo} for the moments of the
$\xi$-distribution whose evolution with energy follows from~\eref{moments}. 
In this formula the number of flavours enters through the running coupling
$\alpha_s(y,n_f)$. The moments evolve at low energy with 3 active flavours 
and with 4 and 5 flavours after passing continuously 
the respective thresholds. As the MLLA is
based on the one-loop expression for $\alpha_s$ there remains a scale
ambiguity. 
The simplest approach would be to put the thresholds at the heavy quark 
masses, i.e., to set the scale of running $\alpha_s$ to $\frac{\sqrt{s}}{2}$. 
However,
let us remind that $\alpha_s$ depends on the transverse momentum 
$k_t$  and kinematics forces 
$k_t \le \frac{1}{4} \frac{\sqrt{s}}{2}$. This would suggest 
to move the thresholds to 4 $m_Q$ or towards even larger values, since the
 effective value of $k_t$ would be in general smaller than the maximum allowed
value.  
Taking the presence of heavier flavours into account in this way, one finds
small deviations at LEP-1 energies in the higher moments $q \ge$ 2~\cite{lo}.
However, for $\xi^*$ using the approximation~\eref{lsdist} the effect of the
higher flavours is only of the order of 1\% and taking $n_f$ = 3 is a very 
good approximation. This is demonstrated in Table~2 which compares the 
predictions for $\xi^*$ under different assumptions on the flavour 
composition and effective thresholds. 

There is also a small difference between the maximum $\xi^*$ of the Limiting
Spectrum~\eref{LS} 
and the modified distribution~\eref{modified} by about 1\%.
This dependence may be considered as systematic uncertainty of the 
predictions. 
 
 Let us also point the attention to the fact that the 
 prediction of the Limiting Spectrum~\eref{lsdist} 
 describes the data quite well without the need of any additional constant
 contribution which might be due to the hadronization process. 
 
\section{Conclusions} 

The analysis of the recent LEP-1.5 data demonstrates that the analytical
perturbative description of inclusive particle distributions in QCD jets 
is in a quite healthy shape. 
The charged 
particle spectrum (up to the overall normalization) and the moments with
$q \ge$ 1 as well as their energy dependence can be described with only one
adjustable scale parameter $\Lambda_{ch}$. 
For these observables, no additional sizeable effects from hadronization are
visible. 
The present data allow one to observe even relatively
small effects caused by higher order terms. Finally, let us emphasize that
in our view it is quite impressive that even a limited sample of hadronic
events (corresponding to integrated luminosity of about 5 $pb^{-1}$) allows 
one to perform so good tests of the perturbative picture of multiple
hadroproduction in jets. Having in mind the prospects of QCD studies 
at LEP-2, this looks rather promising.

\section{Acknowledgements} 

We thank P. Abreu, G. Cowan, L. Del Pozo  and 
G. Dissertori for useful discussions of the experimental data. 
The work was supported in part by the UK Particle Physics and Astronomy
Research Council.

\newpage

\section{Figure Caption} 

\begin{itemize} 

\item[\bf Figure 1.]
{{\bf a.} Charged particle inclusive momentum distributions at $\sqrt{s}$ 
= 91.2 GeV (LEP-1)~\cite{opal} (diamonds) 
and $\sqrt{s}$ = 133 GeV (LEP-1.5)~\cite{opal15} (triangles)
as measured by the 
OPAL Collaboration in comparison with  theoretical predictions of the 
Limiting Spectrum with $\Lambda_{ch}$ = 270 MeV (solid line). Dashed 
lines show the predictions of the Limiting Spectrum after  correction for 
kinematical effects. Theoretical predictions are
also shown for $\sqrt{s}$ = 200 GeV. \\
{\bf b.} Charged particle inclusive momentum distribution at $\sqrt{s}$ 
= 133 GeV (LEP-1.5) measured by ALEPH~\cite{aleph15} (diamonds), 
DELPHI~\cite{delphi15} (squares) and OPAL Collaborations~\cite{opal15} 
(triangles),  and compared to the  same theoretical curves as in {\bf a}.} 

\item[\bf Figure 2.]
{{\bf a.} Maximum of the inclusive momentum distribution $\xi^*$  
as a function of $Y = \log \frac{\sqrt{s}}{2\Lambda_{ch}}$; 
comparison between different 
theoretical predictions: maximum numerically extracted from the shape of the
Limiting Spectrum (solid line), asymptotic formula~\eref{asy} (dashed line), 
eq.~\eref{lsdist} (dotted line). \\
{\bf b.}  Maximum of the inclusive momentum distribution $\xi^*$  
as a function of $Y = \log \frac{\sqrt{s}}{2\Lambda_{ch}}$; 
comparison between experimental
data (see text)  and theoretical prediction 
numerically extracted from the shape of the Limiting Spectrum (solid line); 
$\Lambda_{ch}$ = 270 MeV. 
Crosses are put at $cms$ energies  $\sqrt{s}$ = 200 GeV and 500 GeV.}
\end{itemize}

\newpage 

\section{Table Caption}

\begin{itemize} 

\item[\bf Table 1.]
{The average multiplicity $\bar \N_E$, the average
value $\bar \xi_E$, the dispersion $\sigma^2$, the skewness $s$ and the
kurtosis $k$ of charged particles' energy spectra $E dn/dp$ as a function of 
$\xi_E$ with 
$\Lambda_{ch}$ = 270 MeV extracted from experimental data at $cms$ energies 
$\sqrt{s}$ = 91.2 GeV and 133 GeV. 
In brackets the theoretical predictions of the Limiting Spectrum; 
the second entry in the average multiplicity
column contains the results of the two parameter formula $\bar \N_E = c_1
\frac{4}{9} 2 \bar \N_{LS} + c_2$; the first one the fit with $c_2 = 0$.}

\item[\bf Table 2.]
{Dependence on the number of active flavours 
of the theoretical prediction~\eref{lsdist} for the 
maximum of the inclusive momentum spectrum. 
Results at three different $cms$ energies are shown.}

\end{itemize} 

\newpage


\begin{figure}
          \begin{center}
\mbox{\epsfig{file=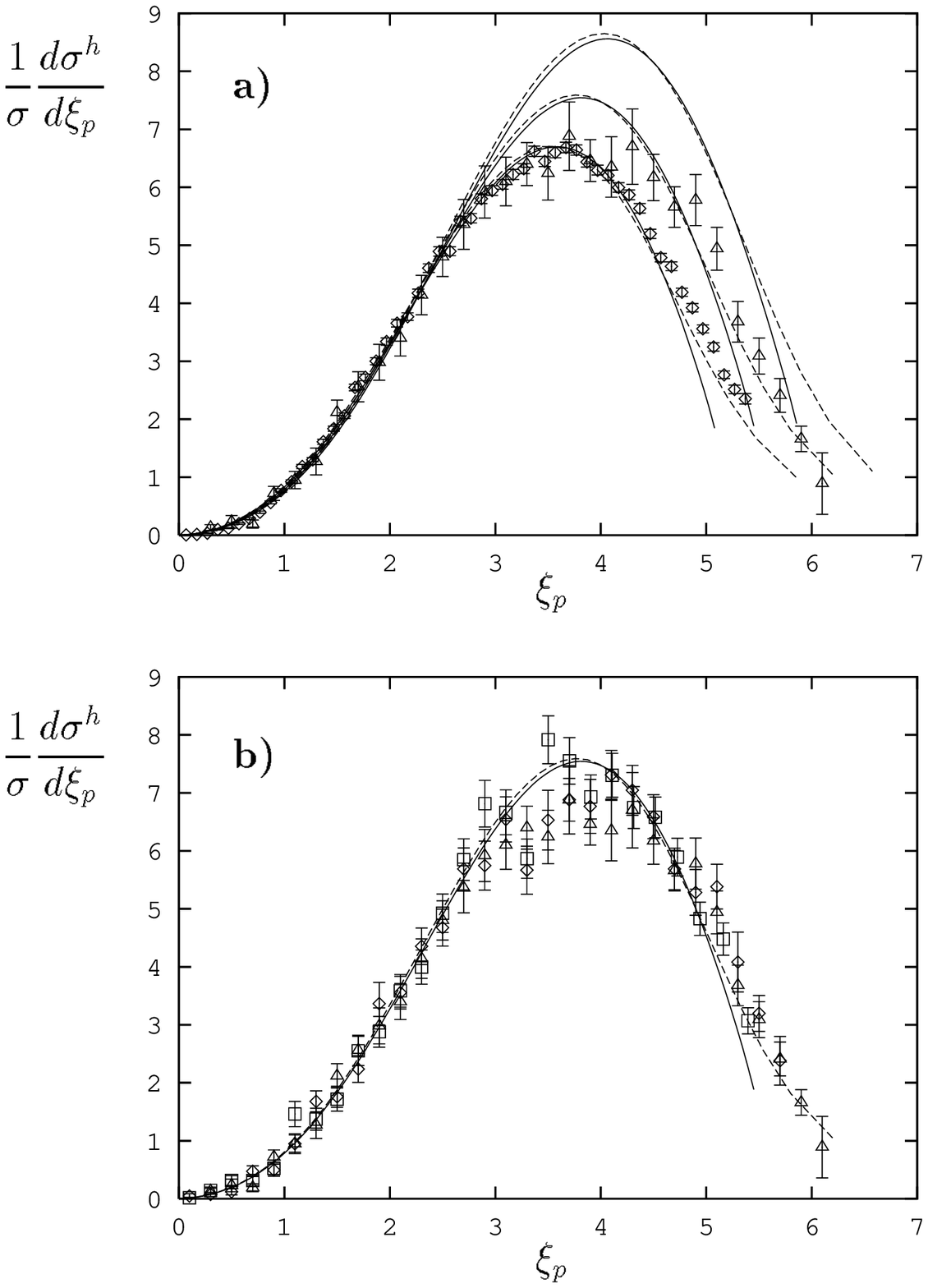,bbllx=2.8cm,bblly=4.cm,%
bburx=20.cm,bbury=26.cm,height=22.cm}}
\end{center} 
\end{figure}
 
\newpage


\begin{figure}
          \begin{center}
\mbox{\epsfig{file=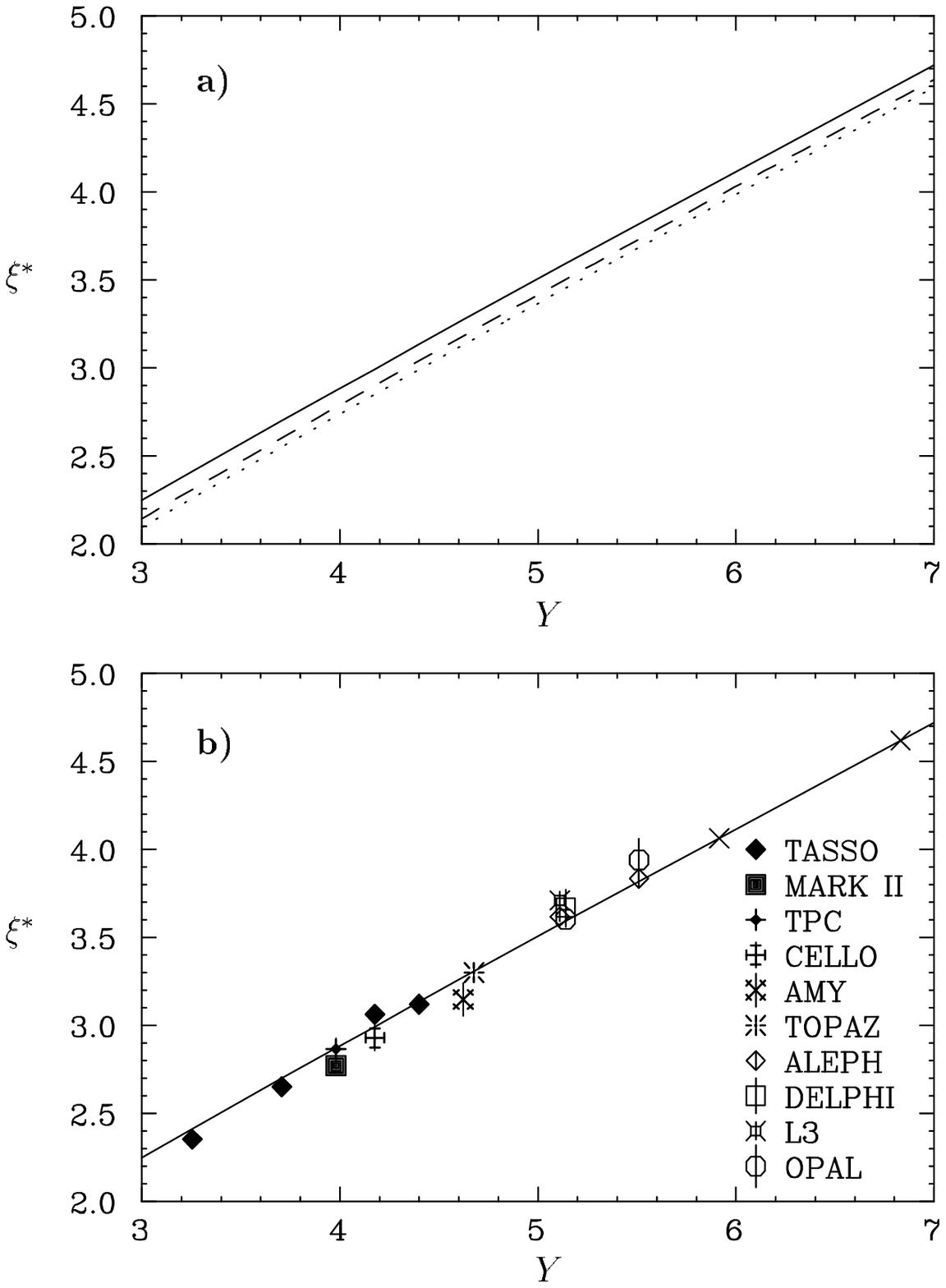,bbllx=2.8cm,bblly=4.cm,%
bburx=20.cm,bbury=26.cm,height=22.cm}}
          \end{center}
\end{figure}

\newpage

\begin{table}     
\vspace{-3.0cm}
\centerline{Table 1}
 \begin{center}
 \vspace{4mm}
 \begin{tabular}{||c|c|c|c|c|c|c||}
  \hline
 Exp.  & $\sqrt{s}$ & 
 $\bar \N_E$ & $\bar \xi_E$ & $\sigma^2$ & $s$ & $k$ \\  
 & (GeV) & & & & & \\ 
  \hline
ALEPH~\cite{aleph}  & 91.2 &
 18.81\pp 1.05 & 
 3.24\pp 0.04 &
 0.99\pp 0.05 & 
 -0.39\pp 0.10 & 
 -0.59\pp 0.32 \\
 DELPHI~\cite{delphi}  & 91.2 &
 19.17\pp 1.00 & 
 3.32\pp 0.02 &
 1.03\pp 0.01 & 
 -0.40\pp 0.02 &
 -0.59\pp 0.07 \\
 L3~\cite{l3}    & 91.2 &
 18.74\pp 1.09 & 
 3.28\pp 0.06 & 
 0.99\pp 0.06 & 
 -0.35\pp 0.13 &
 -0.65\pp 0.40 \\
 OPAL~\cite{opal}  &  91.2 &
 18.95\pp 1.00 & 
 3.29\pp 0.01 & 
 0.99\pp 0.01 & 
 -0.36\pp 0.03 & 
 -0.59\pp 0.09 \\ 
 LEP-1 (avg)         &  91.2 &
 18.93\pp 0.52  & 
 3.29\pp 0.01 & 
 1.01\pp 0.02 & 
 -0.39\pp 0.02 & 
 -0.59\pp 0.05 \\
\hline 
 DELPHI--$\gamma$~\cite{delphi15} & 91.2 &
 19.20\pp 0.20 & 
 3.33\pp 0.04 &
 0.99\pp 0.03 & 
 -0.46\pp 0.05 &
 -0.44\pp 0.18 \\  
 MLLA & 91.2 &  (input,input) & (3.27)  & (1.00)  & (-0.35) &   (-0.52)  \\
\hline 
 ALEPH~\cite{aleph15}   &  133 &
   22.04\pp 0.35   & 
   3.52\pp 0.06  & 
   1.19\pp 0.04 & 
    -0.37\pp 0.07  & 
    -0.62\pp 0.26  \\
 DELPHI~\cite{delphi15}   &    133 &
 22.27\pp 0.31  & 
 3.47\pp 0.05 & 
 1.13\pp 0.05 & 
 -0.40\pp 0.08 & 
 -0.49\pp 0.29 \\
 OPAL~\cite{opal15} &   133 &
 21.50\pp 0.38 & 
 3.51\pp 0.07 & 
 1.19\pp 0.04 & 
 -0.35\pp 0.07 & 
 -0.63\pp 0.25 \\ 
MLLA &   133 &  (22.6,22.5) & (3.50)  & (1.14)  & (-0.34) &   (-0.52)  \\
  \hline
 \end{tabular}
 \end{center}
\label{table}
\end{table}

\newpage

\begin{table}     
\vspace{-3.0cm}
\centerline{Table 2}
 \begin{center}
 \vspace{4mm}
 \begin{tabular}{||c|c|c|c||}
  \hline
 & \multicolumn{3}{c||}{$\xi^*$} \\ 
 \hline
 $n_f$ & $\sqrt{s}$ = 29 GeV & $\sqrt{s}$ = 91 GeV & $\sqrt{s}$ = 133 GeV \\  
  \hline 
 3 & 2.725 & 3.447 & 3.683 \\ 
5 & 2.789 & 3.525 & 3.764 \\  
thresholds at $m_Q$ & 2.746 & 3.482 & 3.721 \\  
thresholds at 4 $m_Q$ & 2.729 & 3.463 & 3.703 \\ 
thresholds at 8 $m_Q$ & 2.726 & 3.455 & 3.695 \\ 
 \hline 
 \end{tabular}
 \end{center}
\label{table2}
\end{table}

\end{document}